\documentclass[aps,prb,preprint,superscriptaddress]{revtex4-1}

\usepackage{amsmath}
\usepackage{rotating}
\usepackage{graphicx}
\usepackage{hyperref}
\usepackage{epstopdf}
\usepackage{color}
\usepackage{booktabs}
\usepackage{longtable}
\usepackage{float}

\usepackage{color}
\usepackage{colortbl}
\usepackage{natbib}
\usepackage{hyperref}

\definecolor{PB}{rgb}{0,0,0}
\definecolor{changes}{rgb}{0.0,0.0,0.0}
\definecolor{changes2}{rgb}{0.0,0.0,0.0}
\definecolor{changes3}{rgb}{0,0,0}
\definecolor{changes5}{rgb}{0,0,0}
\definecolor{changes4}{rgb}{0,0,0}
\definecolor{changes6}{rgb}{0,0,0}
\definecolor{changes7}{rgb}{0,0,0}
\definecolor{lightgray}{rgb}{0.72,0.72,0.72}

\begin{document}


\title{Co(CO)$_n$/Cu(001): Towards understanding chemical control of the Kondo effect}


\author{Marc Philipp Bahlke}
\affiliation{Institut f\"ur Anorganische und Angewandte Chemie, Universit\"at Hamburg, Martin-Luther-King-Platz 6, 20146 Hamburg, Germany}
\author{Peter Wahl}
\affiliation{SUPA, School of Physics and Astronomy, University of St Andrews, North Haugh, St Andrews, Fife KY16 9SS, UK}
\author{Lars Diekh\"{o}ner}
\affiliation{Aalborg University, Department of Materials and Production, Skjernvej 4a, 92220 Aalborg, Denmark}
\author{Carmen Herrmann}
\affiliation{Institut f\"ur Anorganische und Angewandte Chemie, Universit\"at Hamburg, Martin-Luther-King-Platz 6, 20146 Hamburg, Germany}


\date{\today}

\begin{abstract}
The Kondo effect is a many-body phenomenon allowing insight into the electronic and atomistic structure of spin-polarized adsorbates on metal surfaces. Its chemical control is intriguing because it deepens such insight, but the underlying mechanisms are only partly understood. We study the effect of increasing the number of CO ligands attached to a cobalt adatom on copper(001), which correlates with an increase in the Kondo temperature $T_K$ experimentally (P. Wahl et al, Phy. Rev. Lett. 95, 166601 (2005)), by solving an Anderson impurity model parametrized by density functional theory (DFT++). Our results suggest that the orbital responsible for the Kondo effect is $d_{x^2-y^2}$ for the tetracarbonyl, and its combination with $d_{z^2}$ for the dicarbonyl. The molecular structures depend considerably on the approximate exchange--correlation functional, which may be related to the known difficulty of describing CO binding to metal surfaces. These structural variations strongly affect the Kondo properties, which is not only a concern for predictive studies, but also of interest for detecting mechanical deformations and for understanding the effect of tip--adsorbate interactions in the scanning tunneling microscope. Still, by constraining the tetracarbonyl to $C_{4v}$ symmetry, as suggested by experimental data, we find structures compatible with the experimental trend for $T_K$ (employing BLYP-D3+U). This is not possible for the tricarbonyl despite the range of computational parameters scanned. For the tetra- and dicarbonyl, the increased $T_K$ correlates with a larger hybridization function at the Fermi level, which we trace back to an increased interaction of the Co $3d$ orbitals with the ligands.
\end{abstract}

\pacs{}

\maketitle



\section{Introduction}\label{intro}

When a  magnetic atom or molecule is brought in contact with a metallic system, the conduction band electrons can align anti-parallel to the spin direction of the magnetic atom/molecule, due to electron--electron interactions.
If such interaction takes place in impure metallic systems, the resistivity starts to grow with decreasing temperature as opposed to its pure counterpart, which has been first described by J. Kondo in dilute magnetic alloys in 1964 \cite{kondo, kouwenhoven01,tern09}. Since then, the Kondo effect has become a subject of intensive research since it allows insight into fascinating aspects of  electron correlation~\cite{schu17}  and can give information  on the electronic and atomistic structure of molecular adsorbates\cite{zhan15,knaa17,schw18,pacc17,Bazarnik2013,stol18,jaco15b,hira17,Choi2017,kueg18}.

Magnetic molecules on  metallic substrates are of particular interest for spintronics, because of their potential to self-assemble on a surface\cite{Bazarnik2013,Jiang324,zhan15}. The magnetic character of the molecules can be used for storing data, or also for transferring information via spin coupling between different neighboring spin centers \cite{Khajetoorians2011}. For applications in spintronic devices, it is required to understand not only the coupling of the adsorbate's magnetic moment with neighboring spin centers, but also with conduction-band electrons provided by the substrate. As mentioned earlier, the latter can give rise to the Kondo effect, and can be controlled by changing the environment of a spin center. For instance, self-assembly of O$_2$ molecules leads to periodically enhanced Kondo resonances \cite{Jiang324}, and by changing the chain length of Mn$_x$Fe on a CuN$_2$ surface, the Kondo coupling can be controlled \cite{Choi2017}. \\
In this work, we would like to focus on the effect molecular ligands on the Kondo effect~\cite{kara18,zhao05,hein18,tsuk14}. 
Wahl et al. \cite{Wahl2005} have shown that the coupling of the local moment on a Co atom to the conduction band electrons of a Cu(001) substrate can be enhanced by attaching CO ligands to the Co atom (forming Co(CO)$_n$ complexes), by extracting the Kondo temperature from scanning tunneling spectroscopy (STS) experiments. The number of CO ligands might not only affect the hybridization of the Co atom, but will also change the splitting of the $3d$ shell, as a consequence of the different symmetries of the complexes.  For $n$ = 2, a four-fold symmetry was found in the scanning tunneling microscopy (STM), which is due to a thermally induced rotation of the intrinsically two-fold symmetric adsorbate on Cu(001), which happens on a faster time scale than what the STM can resolve. Co(CO)$_3$/Cu(001) is proposed to have no rotational (C$_{3v}$) symmetry, whereas the complex with four ligands exhibits $C_{\mathrm{4v}}$ symmetry (which could result both from an intrinsically $C_{\mathrm{4v}}$-symmetric structure or from a rotating $C_{\mathrm{2v}}$-symmetric one with two opposing ligands being closer to the surface than the other two). The Kondo temperatures $T_{\mathrm{K}}$, as  extracted from a 
Fano fit of the STS spectra, increase with the number of CO ligands:  $165\pm 21$~K ($n$ = 2),  $170\pm 16$~K ($n$ = 3) and $283\pm 36$~K ($n$ = 4) \cite{Wahl2005}. 

We want to gain insight into this behavior as a step towards establishing structure--property relationships for the Kondo effect, by solving the Anderson impurity model parametrized by Kohn--Sham density functional theory (so-called DFT++ methods, following Ref. \cite{lichtenstein98}). In doing so, we will point out how shortcomings in present-day first-principles electronic-structure methods when predicting the atomistic structures of adsorbates on surfaces can strongly affect predicted Kondo properties. This is particularly relevant for systems with CO ligands, since interaction of CO with metal surfaces poses a challenge to electronic structure methods (``CO-puzzle'')\cite{Feibelman2001}.

\section{Methodology}\label{method}
The optimization of molecular structures  on surfaces, in the scope of Kohn--Sham DFT, requires a choice of the approximate exchange--correlation functional that can not only describe the electronic structure of the molecule/adsorbate, but also of the metallic substrate, to predict adsorption distances, angles, adsorption sites and symmetries accurately. 
For molecular adsorbates with their large number of atoms in the unit cell, one often chooses local density approximation (LDA) and generalized-gradient approximation (GGA) type exchange--correlation functionals, as they are  a good compromise between accuracy and computational effort in practice. These classes of exchange--correlation functionals are problematic for the description of the CO ligands, due to an underestimated gap between the highest occupied  and the lowest unoccupied molecular orbital (HOMO and LUMO). This  contributes to the well-known problem of DFT in predicting the correct adsorption sites of CO molecules on different metal substrates, which is known as the ``CO-puzzle''  in the literature \cite{Feibelman2001,Kresse2003, Mason2004,shar08}.

The work of Alaei et al. \cite{Alaei2008} shows that using BLYP (a GGA functional) can at least solve the problem of predicting the incorrect adsorption sites for CO on some metal substrates, such as Rh(111), Pt(111) and Cu(111), but without solving the problem of the underestimated HOMO-LUMO gap of a CO molecule. On the other hand, the work of Favot et al. \cite{Favot2001} shows that PBE (a GGA functional) is able to predict the correct adsorption site of CO on a Cu(001) surface. A more systematic improvement can be reached by taking into account non-local correlation effects \cite{Lazic2010}, as in the scope of the  van der Waals-density functional (vdW-DF with revPBE) developed by Dion et al. \cite{Dion2004}, although its generalizations to spin-polarized systems~\cite{vydr09,Obata2013}
are not broadly available in electronic structure codes. Therefore, we focus on  spin-polarized PBE and BLYP here for structure optimizations, including their DFT+$U$ variants as we have found this to mimick the effect of strong correlation on adsorption distances~\cite{Bahlke2018}. Furthermore, we applied Grimme's dispersion correction (DFT-D3) in all cases \cite{Grimme2010, Grimme2011}.

All Kohn--Sham DFT calculations were performed with the Vienna ab-initio simulation program (VASP) using the projector augmented-wave method \cite{Kresse1996, Kresse1999}. For the carbonyl cobalt complexes under study here, we modeled the Cu(001) surface by a super cell size of 4x4 Cu atoms with five Cu layers in total. Structural relaxation  were done with a 2x2x1 $k$-grid and convergence criteria for the self-consistent field algorithm of 2.7$\cdot$10$^{-5}$~eV, and 0.027~eV/\AA~for the force acting on each atom. In addition to the adsorbates, we allowed the two topmost Cu layers to be relaxed, and also the cell shape (i.e. the lattice parameter). The relaxed lattice parameter  is  3.48-3.49~\AA~for all systems under investigation, consistently for all optimization protocols. In addition, we optimized the systems (adsorbate and the two topmost Cu layers) with a fixed lattice parameter of 3.615~\AA \cite{wyckoff}, which is denoted by the index ``fix''. For DFT+U, we used a on-site Coulomb potential of $U$ = 4.0~eV and $J$ = 0.9~eV applied on the Co $3d$ orbitals.

The electronic structure of the optimized carbonyl cobalt complexes on Cu(001) was then analyzed with a combination of density functional theory and the Anderson impurity model (AIM). For this purpose, we calculated the electronic structure with spin-unpolarized DFT (as usual in the literature) using the PBE exchange--correlation functional and a $k$-grid of 17x17x1 centered around the $\Gamma$-point. A slightly more detailed description of the methodology is given in Ref. [\onlinecite{Bahlke2018}]. The Anderson Hamiltonian reads
\begin{widetext}
\begin{equation}\label{siam_hamilton}
\hat{H}  = \sum_{\nu\sigma}\epsilon_{\nu}\hat{c}^{\dag}_{\nu,\sigma}\hat{c}_{\nu,\sigma} + \sum_{\nu i\sigma} [V_{\nu i}\hat{c}^{\dag}_{\nu,\sigma}\hat{d}_{i,\sigma} + V^*_{\nu i}\hat{d}^{\dag}_{i,\sigma}\hat{c}_{\nu,\sigma}] + 
\sum_{i\sigma}\epsilon_{i}\hat{d}^{\dag}_{i\sigma}\hat{d}_{i\sigma} + \frac{1}{2} \sum_{\substack{ijkl\\\sigma\sigma'}} U_{ijkl} \hat{d}^{\dag}_{i\sigma}\hat{d}^{\dag}_{j\sigma'}\hat{d}_{l\sigma'}\hat{d}_{k\sigma},
\end{equation}
\end{widetext}
where, $\epsilon_i$ is the energy of the $i$th localized $d$ orbital of the impurity (here defined as the Co atom), and $\epsilon_\nu$ is the kinetic energy of the bath electron $\nu$ (in this work, the remainder of the system, i.e. the CO ligands and the copper surface). $\hat{c}_{\nu\sigma}$/$\hat{c}^{\dag}_{\nu\sigma}$ are creation and annihilation operators for electrons with spin $\sigma$ acting on the $\nu$th bath state, whereas $\hat{d}_{i\sigma}$/$\hat{d}_{i\sigma}^{\dag}$ are the corresponding operators acting on the local orbital $i$. The bath electrons are coupled to the impurity via the hybridization $V_{\nu i}$, and $U_{ijkl} = \int drdr'\psi_i^*(r)\psi_j^*(r')\frac{e^2}{|r-r'|}\psi_k(r)\psi_l(r')$ is the local Coulomb interaction (we dropped the spin indices here for simplicity) as introduced by Slater \cite{slater1960}, with $\psi_x$ ($x = i,j,k,l$) being in general any atom-centered basis function. We used  
the parameters ($F^0$, $F^2$ and $F^4$, see below) as derived by Slater for hydrogen-type atomic orbitals. 

 The central quantity extracted from a Kohn--Sham DFT wavefunction is the hybridization $V_{\nu i}$, which we employ to compute an energy-dependent hybridization function $\Delta_{ij}(\omega)$,
\begin{equation}\label{hyb_function}
\Delta_{ij}(\omega) = \sum_{\nu} \frac{V_{\nu i} V_{\nu j}^{*}}{\omega + i0^+ - \epsilon_{\nu}}.
\end{equation}
In practice, this was done by projecting the Kohn--Sham Green's function $G^{KS}_{\nu\nu'}$  onto a set of localized atomic orbitals and then extracting $\Delta_{ij}(\omega)$  from the local non-interacting Green's function $g_{ij}(\omega)$ \footnote{$G^{KS}_{\nu\nu'}$ is the Greens function in the Bloch basis, and since it is obtained from DFT, it can also be regarded as ``non-interacting''. For the non-interacting Green's function in the local basis of the impurity, we use a small letter g.} as
\begin{equation}\label{hyb_function2}
\Delta_{ij}(\omega) =-[g_{ij}^{-1}(\omega) + \epsilon_{ij} - (\omega + 0^+)\delta_{ij}].
\end{equation}
Here, $\epsilon_{ij}$ are the matrix elements of the Kohn--Sham operator in the local basis that has been used in the projection. Diagonalization of the local Kohn--Sham operator leads to the impurity levels $\epsilon_i = \epsilon_{ii}$ introduced in Equation (\ref{siam_hamilton}), and consequently $g_{ij} (\omega) = g_{i} (\omega)$, as well as $\Delta_{ij}(\omega) = \Delta_{i}(\omega)$.

The energy-dependent hybridization function was then used  to solve the Anderson impurity model with the continuous-time quantum Monte Carlo impurity solver in the hybridization expansion.  The imaginary part of $\Delta_{ij}(\omega)$ carries the information on the impurity level being broadened  by the interaction with the bath electrons, while the energy shift caused by this interaction is captured in the  real part of $\Delta_{ij}(\omega)$. The term ``non-interacting'' in this context refers to the DFT solution (although electron correlation effects are captured in the exchange--correlation term), because it is used to parameterize the AIM. The corresponding ``interacting'' solution is subsequently obtained from DFT++.

The Coulomb term of Equation (\ref{siam_hamilton}) was approximated by using only density--density terms (see Ref. \cite{Bahlke2018} for more details), parameterized by the Slater integrals\cite{slater1960, Slater1929} $F^0$, $F^2$, and $F^4$ by using the average Coulomb interaction parameter $U$ ($F^0$) = 4.0~eV and the exchange-interaction parameter $J$ = 0.9~eV ($J = \frac{1}{14}(F^2+F^4) = 0.9$ eV with $\frac{F^4}{F^2} = 0.625$). \\
The correlation energy already captured in the framework of DFT, often called the double-counting (DC) energy,  was corrected by subtracting a term based in  the fully localized limit (FLL) \cite{sawatzky94},
\begin{equation}\label{fll_dc}
\begin{aligned}
E^{\mathrm{FLL}}_{\mathrm{DC}} = \frac{1}{2}UN(N-1) & - \frac{1}{2}JN_{\uparrow}(N_{\uparrow}-1) \\
& - \frac{1}{2}JN_{\downarrow}(N_{\downarrow}-1).
\end{aligned}
\end{equation}
In Equation (\ref{fll_dc}) $N$, $N_{\uparrow}$ and $N_{\downarrow}$ are the total number of electrons, the number of spin-up, and the number of spin-down electrons on the local subspace,  in our case the Co $3d$ shell. It is important to note that the fully localized limit is just an approximation to the real correlation energy already captured in the framework of DFT. Thus, in this work  the DC value was shifted by $\pm$1.0~eV to to evaluate the effect of small errors within the FLL approximation.

Within CT-QMC, we are able to  evaluate the spin--spin correlation function $\chi(\tau)$ on the imaginary time axis 
\begin{equation}\label{spin-spin-cor}
\chi_i(\tau) = \langle \chi_i(\tau) \chi_i(0)\rangle = \langle \hat{S}^i_Z(\tau) \hat{S}^i_Z(0)\rangle.
\end{equation}
In Equation (\ref{spin-spin-cor}), $\hat{S}^i_Z(\tau)$ is the local spin at  imaginary time $\tau$. Evaluating $\chi(\tau)$,  one can gain insight into the magnetic behavior of the system under consideration. The local spin at the initial time $\tau = 0$ is equivalent to the magnetization before interaction with the surroundings takes place. Another special value which is known as the long-time correlation value of $\chi(\tau)$ is $\tau = \beta/2$, from which one can estimate to which extent the impurity electrons are localized on one of the impurity orbitals, or delocalized/screened due to interaction with the substrate and the ligands. The latter case is indicated by a rather rapid drop of $\chi(\tau)$ as $\tau \rightarrow \beta/2)$, whereas a finite value at $\chi(\beta/2)$ suggest that there is spin density located on one of the impurity orbitals even in the presence of the bath.

 \section{Atomistic structure of Co(CO)$_n$ on Cu(001)}\label{struc}
   \begin{figure}[H]
    \centering
    \includegraphics[width=0.35\textwidth]{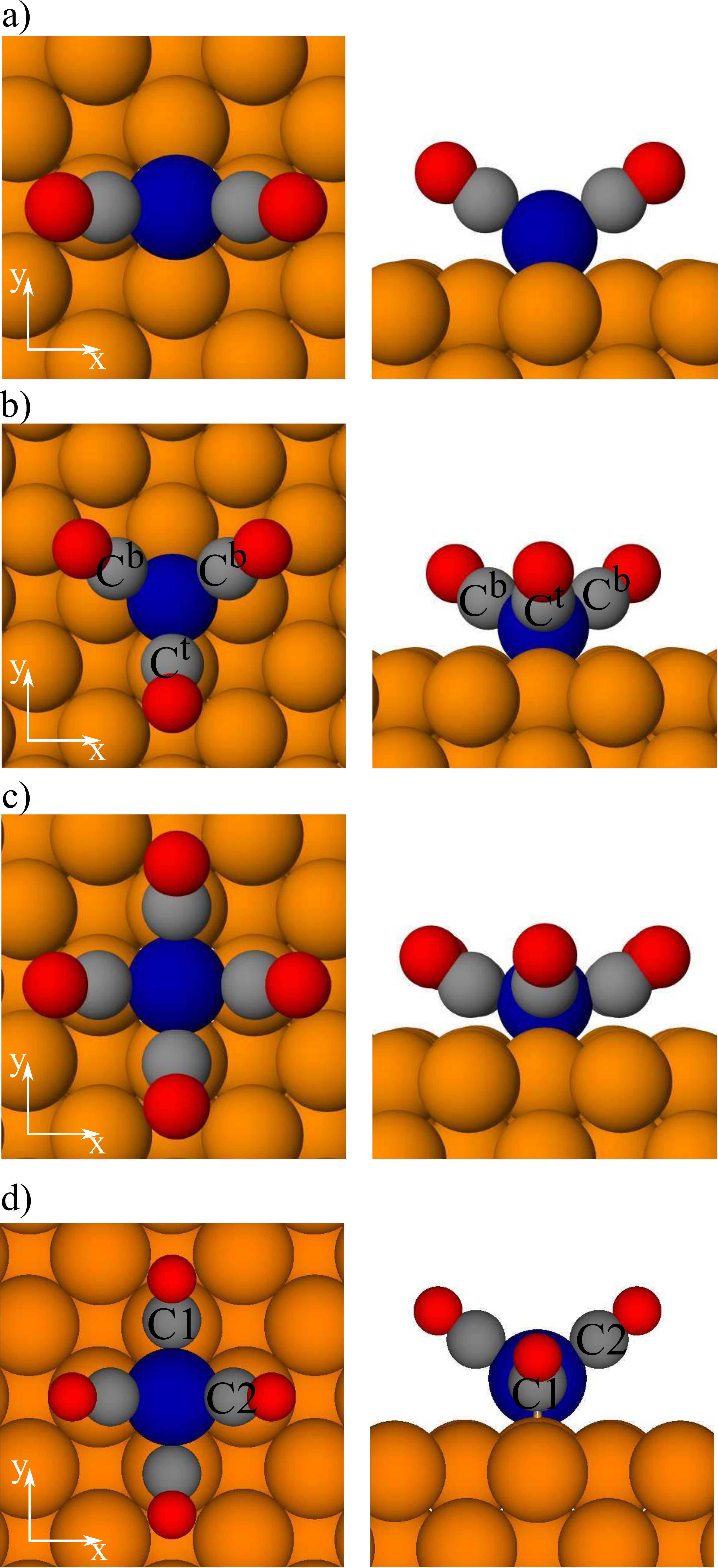}
    \caption{a) Co(CO)$_2$, b) Co(CO)$_3$, c) Co(CO)$_4$ on Cu(100) ($C_{\mathrm{4v}}$) and d) Co(CO)$_4$ on Cu(100) ($C_{\mathrm{2v}}$) as obtained from BLYP-D3+$U$. We labeled the carbon atoms for Co(CO)$_3$ and Co(CO)$_4$ on Cu(100) ($C_{\mathrm{2v}}$) to point out the two symmetrically different CO ligands.}
    \label{carbonyls_opt}
   \end{figure}
 Detailed information on bond lengths, bond angles, and adsorption distances is difficult to access from the experiment. At the same time, as discussed above, metal surfaces with molecular adsorbates in general and with adsorbed CO in particular are challenging for present-day first-principles methods. To address this problem, we have optimized all carbonyl cobalt complexes with PBE-D3 and BLYP-D3, with and without $U$ correction. In some cases we also compared the results of keeping the lattice parameters fixed (for this we use the index ``fix'') with the results obtained by relaxing the cell shape together with the atomic positions of the carbonyl complexes (and the surface atoms of the first two layers, see Section \ref{method} for further details). In the following, we briefly summarize the most important observations about the structures of the carbonyl cobalt complexes on Cu(001), while   detailed information about the structural parameters can be found in the Supplementary Material.
 
 \subsection{Co(CO)$_2$/Cu(001)}
  For Co(CO)$_2$/Cu(001), one consistently obtains $C_{2v}$ symmetry for all exchange--correlation functionals mentioned above, which is in agreement with  Ref. [\onlinecite{Wahl2005}]. It can also be confirmed that the rotational barrier of Co(CO)$_2$/Cu(001) is rather low (6.4~kJ/mol = 66~meV as obtained from PBE$_{\mathrm{fix}}$-D3), which would support the observation of Wahl et al. \cite{Wahl2005} that this molecule is rotating faster than the STM time resolution on Cu(001) in the experimentally accessible temperature range. 
  
  Given that present-day approximate DFT has problems to describe the Kondo screening of the magnetic moment, we would expect to obtain a nonzero magnetic moment from spin-polarized DFT optimizations for systems showing a Kondo effect. 
  However, we do not find such a  magnetic moment on the adsorbate when using PBE-D3 and BLYP-D3, both for the optimizations including and excluding the cell shape. When we apply a Hubbard $U$ correction on the Co atom for BLYP-D3, we do obtain a magnetic moment of  1.0 $\mu_\mathrm{B}$ on the Co atom of Co(CO)$_2$/Cu(001) when we include the cell shape in the optimization. This increased spin localization is in line with the tendency of DFT+U to correct for electron overdelocalization~\cite{kuli15}. As shown in Table \ref{adsorptiondistances}, the BLYP-D3+U spin localization has consequences for the adsorption distance  as indicated by the increased Co--surface distance $d_{\mathrm{Co-surf.}}$ compared with the BLYP-D3-optimized structure. 
  A similar increase of the adsorption distance can be observed as a result of cell shape relaxation, as suggested by comparing the values for BLYP-D3$_{\mathrm{fix}}$  (1.39~\AA) and BLYP-D3 (1.60~\AA). This might come from the slightly reduced lattice parameter (see Section \ref{method}) in cases with cell shape optimization, which effectively decreases the size of the four-fold hollow position and thus increases the adsorbate--surface distance. We will later show that there is a delicate dependence on the Kondo properties  obtained from DFT++ on the adsorption distance. 
  
   \subsection{Co(CO)$_3$/Cu(001)}
   The DFT-predicted symmetry of Co(CO)$_3$/Cu(001) in this study differs from the one reported in Ref. \onlinecite{Wahl2005} for all exchange--correlation functionals under study.  Here, we obtain a rather $C_{3v}$-like\footnote{It is actually not a perfect $C_{\mathrm{3v}}$ symmetry, because the two CO ligands in a bridged position (with respect to the Cu(001) surface) have slightly different structural parameters, compared to the one in top position.} symmetry of the molecule  (see Fig. \ref{carbonyls_opt}), whereas in Ref. \onlinecite{Wahl2005},  both the experimental and the  reported DFT structure suggest no such rotational symmetry. Of course one may debate whether the STM results (Fig. 1 in Ref. [\onlinecite{Wahl2005}]) clearly exclude $C_{3v}$ symmetry. However, since our optimized structures do not appear consistent with these data, and since we could not find Kondo features in the DFT++ data for any of these structures, we assume that the range of DFT variants employed here is not capable of describing the tricarbonyl structure reliably, and will not focus on it any further within the scope of this work (see Supplementary Material for more details).


   \begin{table}[h]
      \caption{Optimized adsorption distances of the carbonyl cobalt complexes on Cu(001) in~\AA~as obtained  from different DFT protocols. The index ``fix'' is used in cases where the experimental lattice constant was used. 
}
     \centering
     \begin{tabular}{c c c c c  c}
    \toprule[1.5pt]
    		& PBE-D3$_{\mathrm{fix}}$ \ \	& PBE-D3 \ \	& BLYP-D3$_{\mathrm{fix}}$ \ \	& BLYP-D3 \ \	&  BLYP-D3$+U$\\
    \midrule[0.75pt]
    Co (isolated)   &   1.46	 &  -   &  -	 &  -   &  1.78	\\
    Co(CO)$_2$   &   1.33	 &  1.54   &  1.39	 &  1.60   &  1.74	\\
    Co(CO)$_3$   &   1.58  &  -  &   -  &  1.75  &  1.76   \\
    Co(CO)$_4$ $(C_{\mathrm{2v}})$  &   1.90  &  -   &  -   &  2.07  &  2.26   \\  
    Co(CO)$_4$ $(C_{\mathrm{4v}})$  &   1.69  &  -   &  -   &  1.89  &  1.85   \\
    \bottomrule[1.5pt]
     
     \end{tabular}
     \label{adsorptiondistances}
     \end{table}
     
      \subsection{Co(CO)$_4$/Cu(001)}
 In the case of Co(CO)$_4$/Cu(001), it is challenging to obtain a four-fold ($C_{4v}$)-symmetric structure as suggested by the STM experiments of Wahl et al. \cite{Wahl2005}, because in all attempts to optimize the structure, the two-fold ($C_{2v}$) symmetry was found to be lower in energy by about 30.2 kJ/mol (PBE$_{\mathrm{fix}}$-D3) to 42.1 kJ/mol (BLYP-D3$+U$). The vdw-DF functional (revPBE$_{\mathrm{fix}}$) predicts $C_{2v}$ symmetry to be 70.0~kJ/mol lower in energy than $C_{4v}$ symmetry. The rotational barrier of the molecule (as obtained for the PBE$_{\mathrm{fix}}$-D3 structure) in $C_{2v}$ symmetry is 20~kJ/mol (0.21~eV), which suggests  that a hypothetical rotation of this molecule in the STM experiments by Wahl et al. is not responsible for the observed $C_{4v}$ symmetry. 
 In the experiment, the preparation of the carbonyl complexes  was done by first depositing cobalt on Cu(001), and then saturating the surface with CO molecules. DFT (BLYP+$U$) suggests a rather short adsorption distance (Table \ref{adsorptiondistances}, also compare Ref.~\cite{Huang2008}) of an isolated Co on Cu(001) (1.78~\AA), which is closer to the tetracarbonyl in $C_{4v}$ symmetry (1.85~\AA) than to the one in $C_{2v}$ symmetry (2.26~\AA). It is conceivable that the formation of the $C_{4v}$-symmetric system is kinetically favored, due to the adsorption distance being closer to that of an isolated Co on Cu(001). Although the sample was annealed to T = 200~K-300~K, the formation of the probably more stable $C_{2v}$ structure (as suggested by DFT) may be inhibited.
 
 In the DFT and DFT+$U$ calculations, we again do not obtain a local magnetic moment on  Co(CO)$_4$/Cu(001) in both symmetries.  This could make  an  interpretation in terms of a Kondo effect difficult, but as we will see later, it is indeed possible to identify features in the DFT++ electronic structure for Co(CO)$_4$/Cu(001) in  $C_{4v}$ symmetry which are in agreement with the experimentally observed Kondo effect. For the corresponding $C_{\mathrm{2v}}$ symmetry, no Kondo properties could be found, suggesting that only the $C_{4v}$ symmetry is consistent with the experimental observations. 
 

%
%
\section{Kondo properties of Co(CO)$_n$/Cu(001)}\label{sec:kondo}

In the following, we will focus on the structures optimized with BLYP-D3+$U$ (with optimization of the cell shape), since 
these show a nonzero local moment on the Co atom for the dicarbonyl complex. For  Co(CO)$_4$ on Cu(001), we limit the discussion to the $C_{\mathrm{4v}}$ structure, because it fits best to the experimentally observed symmetry (see Figure 1 of Ref. [\onlinecite{Wahl2005}]), and allows for an interpretation in terms of a Kondo effect (see below).
For these structures, we  parametrize the AIM with closed-shell PBE, in order to correctly describe the non-magnetic character of the Cu(001) surface (and thus the coupling of the Co $3d$ orbitals with a non-magnetic metal).

As the Kondo effect manifests itself as a sharp feature in the spectral function  at the Fermi energy, we aim at identifying the relevant Co $3d$ orbitals that might contribute to the experimentally observed zero-bias anomaly from the spectral functions of the individual orbitals\cite{Wahl2005}.  Nozi\`eres \cite{nozi74} showed that a Kondo effect can be described within the Fermi liquid theory. We will use this to identify the transition point to the Fermi liquid regime, as an approximated value for the Kondo temperature $T_{\mathrm{K}}$. Therefore, we will analyze the temperature dependence of the spin--spin correlation function $\chi(\tau)$ at the special value $\tau$ = $\beta/2$, which should behave as $T^2$ in the Fermi liquid regime. Furthermore, we will investigate  the so-called first Matsubara-frequency rule \cite{Chubukov2012}, which states that Im$\Sigma(\omega_0$) should go linearly to zero as $T\rightarrow 0$~K, as another tool for probing the Fermi liquid properties of the di- and tetracarbonyl systems. 

\subsection{Spectral properties of Co(CO)$_n$/Cu(001)}
 \begin{figure}[H]
  \centering
  \includegraphics[width=0.7\textwidth]{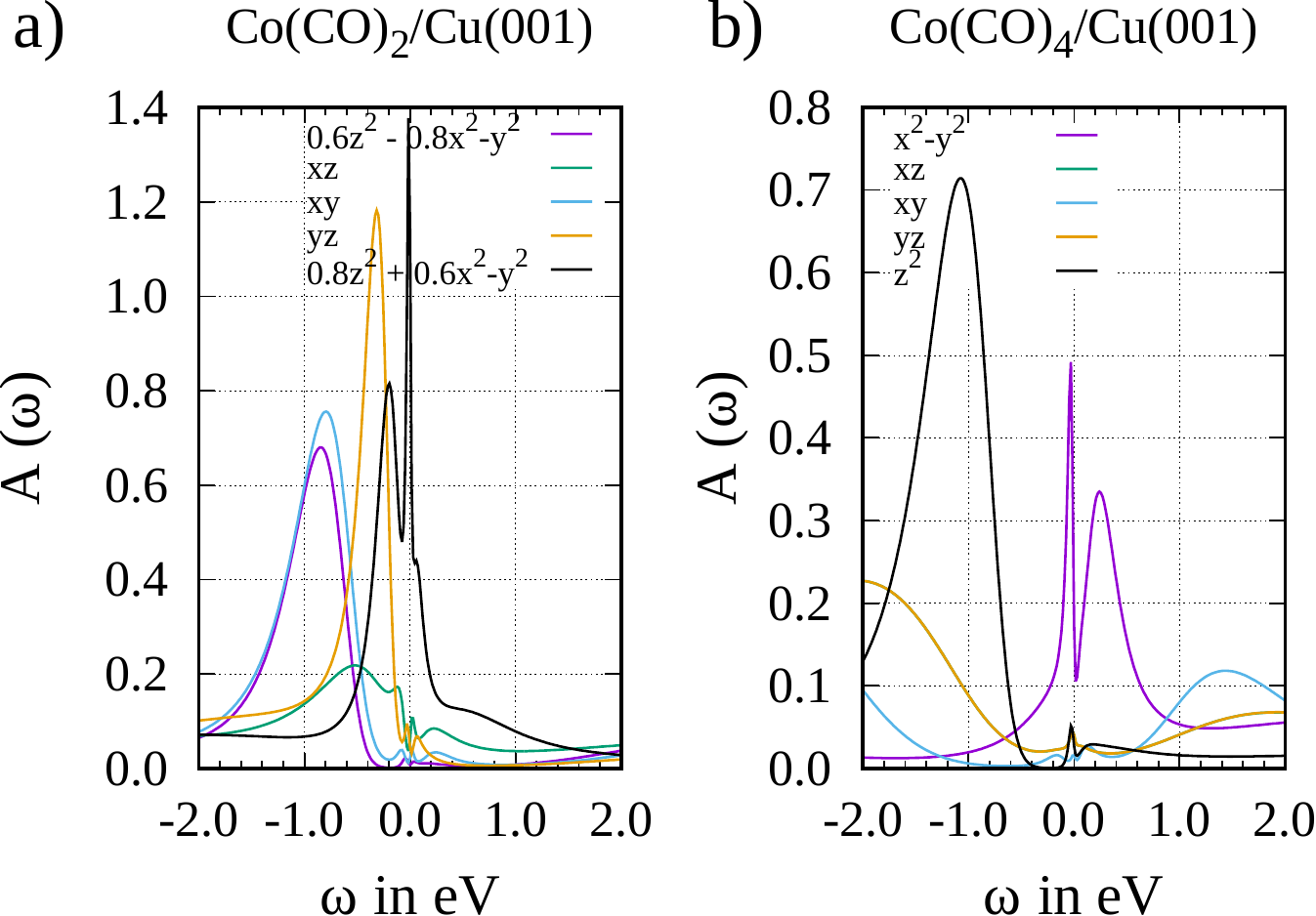}
  \caption{ Spectral functions as obtained from PBE++  for  a) Co(CO)$_2$ and  b) Co(CO)$_4$ on Cu(100) ($C_{\mathrm{4v}}$) at $\beta$ = 100~eV$^{-1}$ ($T$ = 116~K). Here the fully localized limit was used with $U$ = 4.0~eV and $J$ = 0.9~eV.}
  \label{carbonyls_spectra}
 \end{figure}
In Figure \ref{carbonyls_spectra}, all Co $3d$ spectral functions of Co(CO)$_2$ and  Co(CO)$_4$ on Cu(100) ($C_{\mathrm{4v}}$) are shown at $T$ = 116~K. Due to diagonalization of the Co $3d$ subspace, some of the Co $3d$ orbitals are mixed.  Co(CO)$_2$/Cu(001) (Figure \ref{carbonyls_spectra} a)) shows a sharp feature at the Fermi energy ($\omega$ = 0.0~eV) for the Co $3d_{\mathrm{0.8z^2 + 0.6x^2-y^2}}$ orbital, whereas the remaining Co $3d$ orbitals only have broader features below the Fermi energy (in the energy range shown here). The sharp feature of this orbital  thus makes it a promising candidate for causing the experimentally observed Kondo effect (which will be further investigated  in terms of its Fermi liquid properties later on). This is an interesting contrast to the bare Co atom on Cu(001), where (at least at relatively large temperature) it is likely the $d_{z^2}$ orbital which is causing the Kondo properties\footnote{It has been argued that a second orbital may be contributing to the Kondo properties of Co/Cu(001) at temperatures too low to be reached with the methodology employed here~\cite{jaco15,baru15}}.


In the case of Co(CO)$_4$/Cu(001) in $C_{\mathrm{4v}}$ symmetry, we would like to re-emphasize that  DFT (using GGA-type functionals) predicts a magnetic moment of 0.0$\mu_{\mathrm{B}}$ on the Co atom. Nonetheless, the spectral function of Co(CO)$_4$/Cu(001) obtained from DFT++ ( Figure \ref{carbonyls_spectra} b)) shows a sharp feature at the Fermi energy for the Co $3d_{\mathrm{z^2}}$ orbital. Based on this observation, there is reason to believe that DFT++ predicts a finite local moment on the Co atom of Co(CO)$_4$/Cu(001) in this orbital, which is probably screened due to a Kondo effect and thus leading to the sharp feature in the spectral function, as we will confirm later. Due to the $C_{\mathrm{4v}}$ symmetry, Co(CO)$_4$/Cu(001) would be a promising candidate for a so-called orbital Kondo effect (similar as reported for cobalt-benzene sandwich molecules \cite{Karolak2011} and for Co adatoms on graphene~\cite{wehl08}), because it has two degenerate orbitals ($d_{\mathrm{xz/yz}}$). However, the lack of a resonance at the Fermi energy for these orbitals suggests that this is not the case.

To conclude, for Co(CO)$_2$/Cu(001) and Co(CO)$_4$/Cu(001) ($C_{\mathrm{4v}}$), we could identify a sharp, Kondo-like feature at the Fermi energy, which gives us the opportunity to investigate the increasing Kondo temperature with an increasing number of CO ligands, for answering the question of how the Kondo effect can be chemically controlled. 

\subsection{Fermi liquid properties of Co(CO)$_n$/Cu(001)}
We  study the Fermi liquid properties to probe the existence of a Kondo effect as suggested by the spectral functions for Co(CO)$_2$ and for Co(CO)$_4$ in $C_{\mathrm{4v}}$ symmetry. 

\subsubsection{Spin--spin correlation function at high temperatures: Is there a magnetic moment to be screened?}
\begin{figure}[H]
  \centering
  \includegraphics[width=1\textwidth]{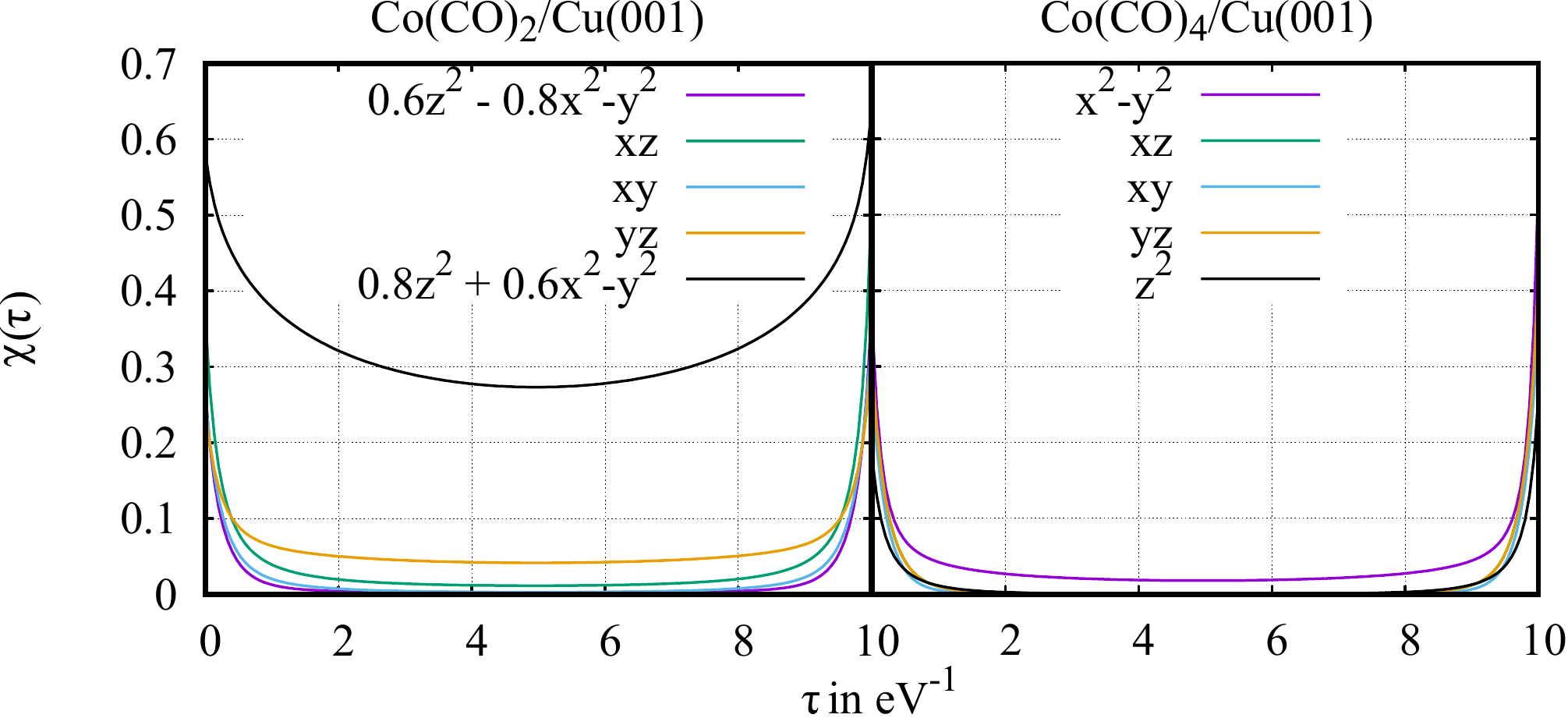}
  \caption{Spin-spin correlation function $\chi(\tau)$ as obtained from PBE++ at $\beta$ = 10~eV$^{-1}$ ($T$ = 1160~K) of all carbonyl cobalt complexes on Cu(001) under consideration here. The PBE++ calculation was done with $U$ = 4.0~eV, $J$ = 0.9~eV, and using the fully localized limit for estimation the double-counting correction.}
  \label{carbonyls_spincor}
 \end{figure}

First, we consider the spin--spin-correlation functions of the carbonyl cobalt complexes in Figure \ref{carbonyls_spincor}. Here,  $\chi(\tau)$ is shown at $T$ = 1160~K, because we expect that at this temperature, no Kondo screening takes place, and a finite local moment should be observed. For Co(CO)$_2$/Cu(001), $\chi(\tau)$ of the $3d_{\mathrm{0.8z^2 + 0.6x^2-y^2}}$ and  $3d_{\mathrm{yz}}$ orbitals drops to a non-zero value as $\tau\rightarrow \beta/2$, suggesting a persisting spin-density in these orbitals. Most of the magnetic moment observed for  Co(CO)$_2$/Cu(001) is located in the Co $3d_{\mathrm{0.8z^2 + 0.6x^2-y^2}}$ orbital, as suggested by the larger value at $\chi(\beta/2)$. This observation suggests that the feature at $\omega = 0.0$~eV in the spectral function of the Co $3d_{\mathrm{0.8z^2 + 0.6x^2-y^2}}$ orbital is indeed a signature of a Kondo effect.

For Co(CO)$_4$/Cu(001) ($C_{\mathrm{4v}}$), only the Co $3d_{\mathrm{x^2-y^2}}$ orbital shows a finite value of $\chi(\beta/2)$ at $T$ = 1160~K (Figure \ref{carbonyls_spincor}), which is the same orbital contributing spectral weight at $\omega = 0.0$~eV in form of a sharp feature. Thus, this orbital is a promising candidate for causing the observed Kondo effect in Co(CO)$_4$/Cu(001). For the $C_{\mathrm{2v}}$ symmetry, we observe that $\chi(\beta/2)$ drops to zero for all Co $3d$ orbitals, similar as for Co(CO)$_3$/Cu(001) (see Supplementary Material).

To summarize, we were able to identify promising candidates for Kondo-relevant orbitals for Co(CO)$_2$/Cu(001) ($3d_{\mathrm{0.8z^2 + 0.6x^2-y^2}}$) and Co(CO)$_4$/Cu(001) in $C_{\mathrm{4v}}$ symmetry ($3d_{\mathrm{x^2-y^2}}$) from the spectral and spin--spin correlation functions. 



\subsubsection{Temperature dependence of $\chi(\beta/2)$}
  \begin{figure}[H]
  \centering
  \includegraphics[width=1.0\textwidth]{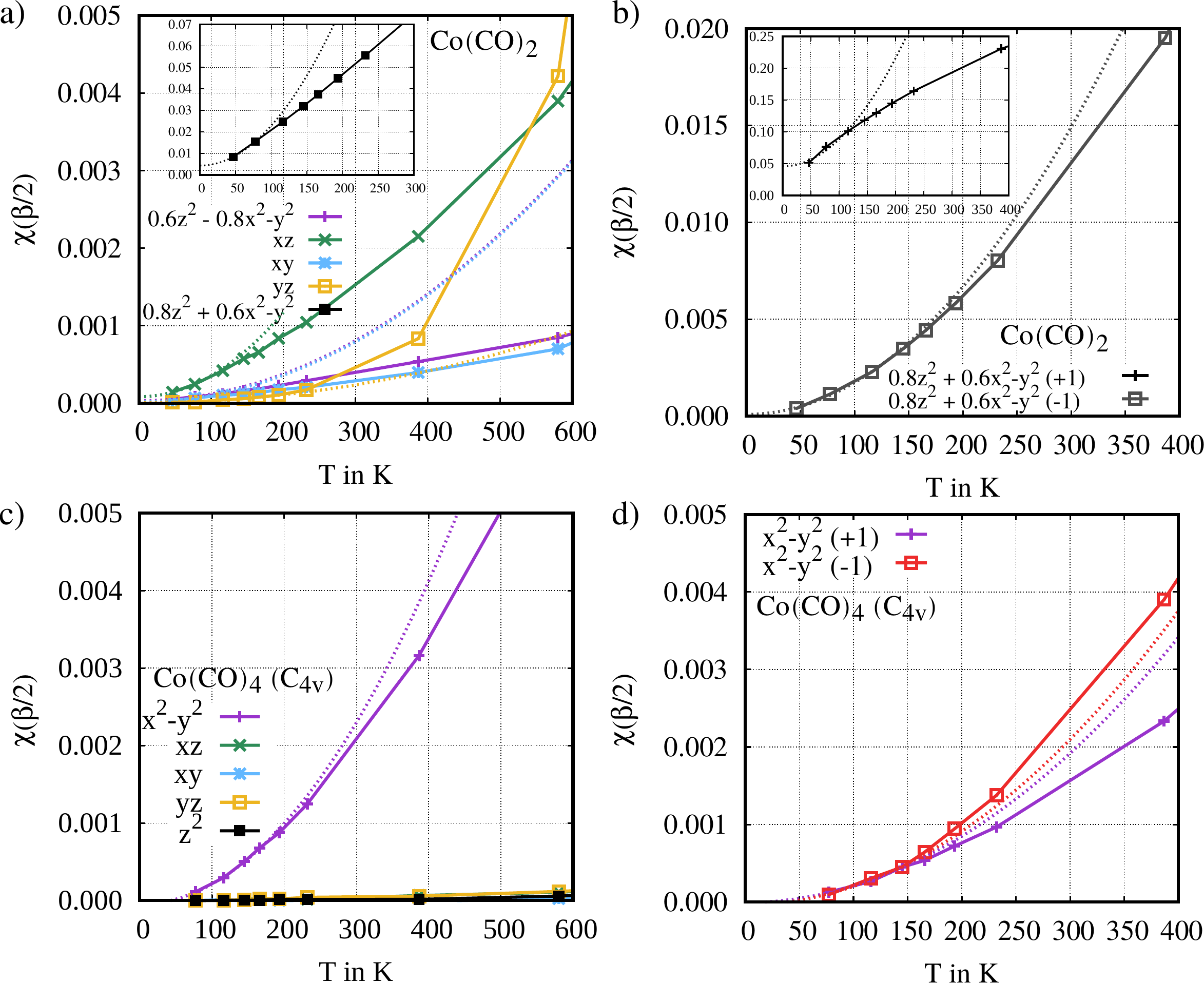}
  \caption{a) $\chi(\beta/2)$ as a function of $T$ for all Co $3d$ orbitals of Co(CO)$_2$/Cu(001) using the fully localized limit for the double-counting correction. b) $\chi(\beta/2)$ as a function of $T$ for the Co $3d_{\mathrm{0.8z^2 + 0.6x^2-y^2}}$ orbital of Co(CO)$_2$/Cu(001).  Here the value of the double-counting was shifted by $\pm$1.0~eV with respect to the FLL value. c) $\chi(\beta/2)$ as a function of $T$ for all Co $3d$ orbitals of Co(CO)$_4$/Cu(001) (C$_{\mathrm{4v}}$) using the fully localized limit for the double-counting correction. b) $\chi(\beta/2)$ as a function of $T$ for the Co $3d_{\mathrm{x^2-y^2}}$ orbital of Co(CO)$_4$/Cu(001) (C$_{\mathrm{4v}}$). Again, the value of the double-counting was shifted by $\pm$1.0~eV with respect to the FLL value.
  The fits  are quadratic fits of the first two data points. Note that the fits are meant as a guide for the eye to check whether the Fermi-liquid behavior is fulfilled or not.}
  \label{chibeta_di_tetra}
 \end{figure}
 
\begin{figure}[H]
    \centering
    \includegraphics[width=1.0\textwidth]{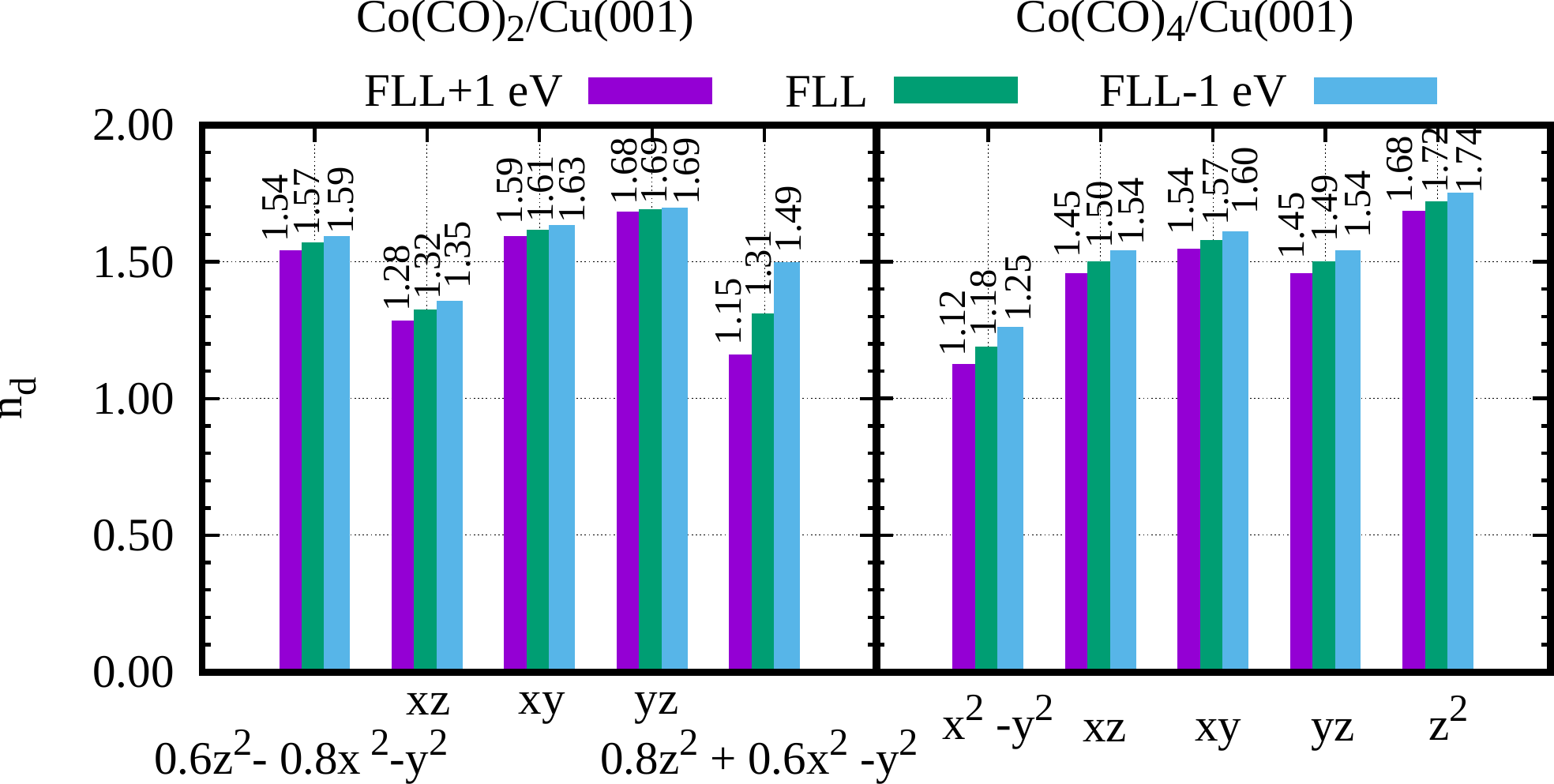}
    \caption{Co $3d$ fillings of Co(CO)$_2$ and Co(CO)$_4$ on Cu(001) as obtained from PBE++ at $\beta$ = 100~eV$^{-1}$~eV for different values of the FLL double-counting correction. ``FLL'' is the value as obtained from Equation (\ref{fll_dc}). }
    \label{carbonyls_fillings}
   \end{figure}

As mentioned at the beginning of Sec. \ref{sec:kondo}, the Fermi liquid behavior is manifested as a $T^2$ dependence of $\chi(\beta/2)$. In Figure \ref{chibeta_di_tetra} a), we study this behavior for all Co $3d$ orbitals of CO(CO)$_2$/Cu(001) as obtained within the fully localized limit.  In all cases  $\chi(\beta/2)$ drops as $T$ is lowered, presumably approaching zero as $T\rightarrow 0~K$. For the Kondo-relevant orbital $3d_{\mathrm{0.8z^2 + 0.6x^2-y^2}}$, one does not observe $T^2$ dependence (and thus no Fermi liquid behavior), due to the non-zero intercept with the ordinate. A better agreement within a Fermi liquid behavior  can be observed by shifting the DC value by $-1.0$~eV  from the original FLL value (Figure \ref{chibeta_di_tetra} b)), which increases the  occupation of the $3d_{\mathrm{0.8z^2 + 0.6x^2-y^2}}$ orbital from 1.31 (FLL) to 1.49 electrons (see Figure \ref{carbonyls_fillings}). 

Here one observes that the $3d_{\mathrm{0.8z^2 + 0.6x^2-y^2}}$ orbital of Co(CO)$_2$/Cu(001) behaves quadratically in $T$ below $T = 150$~K. Shifting the DC value by $+1.0$~eV from the original FLL value, the filling in the $3d_{\mathrm{0.8z^2 + 0.6x^2-y^2}}$ orbital reduces to 1.15 electrons, and  the values of $\chi(\beta/2)$ are significantly increased as compared to the results obtained within the FLL. The  sensitivity of $\chi(\beta/2)$ to the shift of the DC value comes from the fact that the occupation on the $3d_{\mathrm{0.8z^2 + 0.6x^2-y^2}}$ orbital of Co(CO)$_2$/Cu(001) changes from 1.15 (FLL$+1$~eV) to 1.49 (FLL$-1$~eV), as shown in Figure \ref{carbonyls_fillings}. This indicates that the charge fluctuations in this orbital are strongly increased when the DC value is shifted by $-1$~eV. This leads to the observed transition to the Fermi liquid regime at roughly $T$ = 150~K, whereas for the other fillings under consideration here, it is not possible to see this transition for the temperature range considered.

In Figure \ref{chibeta_di_tetra} c), we depict $\chi(\beta/2)$ as a function of $T$ for Co(CO)$_4$/Cu(001) ($C_{\mathrm{4v}}$) as obtained within the fully localized limit. $\chi(\beta/2)$ of the Kondo-relevant orbital ($3d_{\mathrm{x^2-y^2}}$) behaves as $T^2$ at temperatures below $T$ = 165~K, pointing to a transition to the Fermi liquid regime at this temperature. Shifting the DC by $\pm 1$~eV with respect to the FLL value alters the temperature at which the transition to the Fermi liquid regime is observed only little (see Figure \ref{chibeta_di_tetra} d)). As one can see from Figure \ref{carbonyls_fillings}, this might be due to the fact that the filling on the $3d_{\mathrm{x^2-y^2}}$ orbital (1.12 (FLL$+1$~eV) to 1.25 (FLL$-1$~eV) electrons) is not affected as strongly as in the case of the $3d_{\mathrm{0.8z^2 + 0.6x^2-y^2}}$ orbital in Co(CO)$_2$/Cu(001).

\subsubsection{First Matsubara frequency rule}
\begin{figure}[H]
  \centering
  \includegraphics[width=1.0\textwidth]{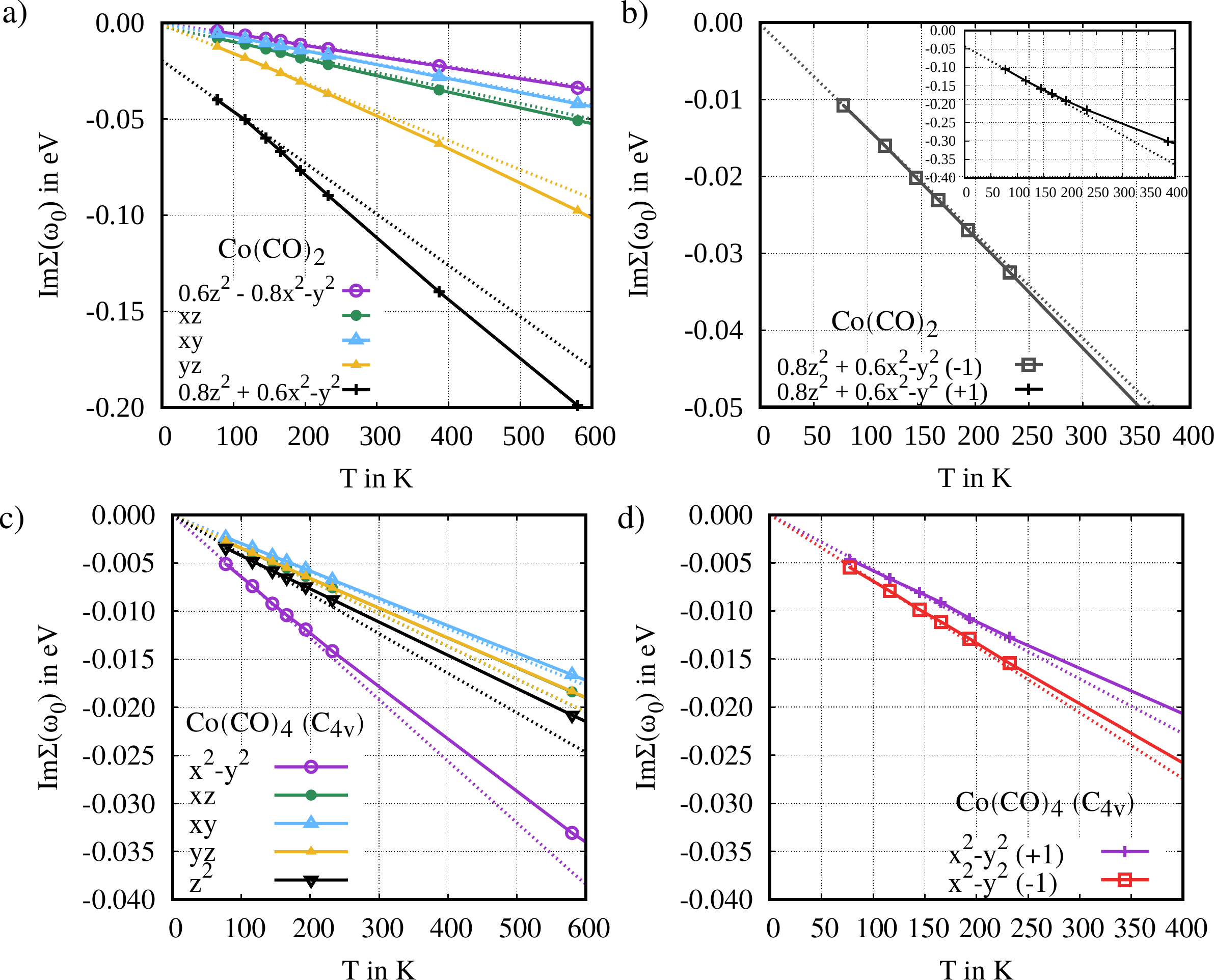}
  \caption{a) Im$\Sigma(\omega_0)$ as a function of $T$ for all Co $3d$ orbitals of Co(CO)$_2$/Cu(001) using the fully localized limit. b) Im$\Sigma(\omega_0)$ as a function of $T$ for the Co $3d_{\mathrm{0.8z^2 + 0.6x^2-y^2}}$ orbital of Co(CO)$_2$/Cu(001). Here the value of the double-counting was shifted by $\pm$1.0~eV with respect to the FLL value. c) Im$\Sigma(\omega_0)$ as a function of $T$ for all Co $3d$ orbitals of Co(CO)$_4$/Cu(001) (C$_{\mathrm{4v}}$) using the fully localized limit. d) Im$\Sigma(\omega_0)$ as a function of $T$ for the Co $3d_{\mathrm{x^2-y^2}}$ orbital of Co(CO)$_4$/Cu(001) (C$_{\mathrm{4v}}$). The fits in a)-d) are linear fits as a guide for the eye to check whether the Fermi-liquid behavior is fulfilled or not.}
  \label{matsu_di_tetra}
 \end{figure}

We find that a similar behavior of Co(CO)$_2$/Cu(001) concerning the Fermi liquid properties can be found by considering the so-called first Matsubara-frequency rule depicted in Figure \ref{matsu_di_tetra} a) and b). The only agreement with Fermi liquid behavior can be observed if the DC value is shifted by $-1.0$~eV, as indicated by the linear behavior of Im$\Sigma(\omega_0)$ as $T \rightarrow 0$~K for temperatures below $T$ = 150~K. Thus, both $\chi(\beta/2)$ and Im$\Sigma(\omega_0)$ as a function of $T$ conclude that the $3d_{\mathrm{0.8z^2 + 0.6x^2-y^2}}$ orbital behaves as a Fermi liquid, but the transition temperature strongly depends on the Co $3d$ filling. Nevertheless, this gives further support to our initial conclusion that the sharp feature observed in the spectral function (Figure \ref{carbonyls_spectra} a)) is a signature of a Kondo effect.

 For Co(CO)$_4$/Cu(001), the first Matsubara-frequency rule (Figure \ref{matsu_di_tetra} c) and d)) is also fulfilled for the $3d_{\mathrm{x^2-y^2}}$ orbital at all DC values under consideration here, and affirms furthermore the transition to the Fermi liquid regime roughly below $T$ = 165~K. This gives reason to believe that the observed feature at $\omega = 0.0$~eV in the spectral function is a true Kondo signature. 

 \subsubsection{Discussion}
 
To summarize, we could show that the suggested Kondo-relevant orbitals of Co(CO)$_2$/Cu(001) and Co(CO)$_4$/Cu(001) ($C_{\mathrm{4v}}$) display the Fermi liquid properties expected for a Kondo system. However, for Co(CO)$_2$/Cu(001), the transition temperature, or Kondo temperature, depends delicately on the choice of the double-counting value. Within this work, it is only possible to see a transition to the Fermi liquid regime  (at $T$ = 150~K) if the filling on the Co $3d_{\mathrm{0.8z^2 + 0.6x^2-y^2}}$ orbital  is increased by shifting the DC value by $-1$~eV from the FLL value. In the case of Co(CO)$_4$/Cu(001) ($C_{\mathrm{4v}}$), the transition to the Fermi-liquid regime is more robust against changes of the double-counting value, as confirmed unanimously by $\chi(\beta/2)$ and Im$\Sigma(\omega_0)$ as a function of $T$. We find that for all DC values the transition to the Fermi-liquid regime of Co(CO)$_4$/Cu(001) is at roughly $T$ = 165~K.

Concerning  the question of how the Kondo effect is controlled by the number of CO ligands, one is now faced with the problem that the exact double-counting correction for both systems is unknown, and that the Kondo temperature of  Co(CO)$_2$/Cu(001) is only detectable (within the electronic temperatures reached here) if the DC correction is shifted towards smaller values. However, comparing the Fermi liquid behavior of both systems as obtained from the fully localized limit 
Co(CO)$_4$/Cu(001) has indeed a larger Kondo temperature than Co(CO)$_2$/Cu(001) in qualitative agreement with
the experimental data, since in the latter case the transition to the Fermi liquid regime occurs at temperatures lower than the ones considered here. Later, we will give an estimation of $T_{\mathrm{K}}$ based on the hybridization function  of the Co $3d_{\mathrm{0.8z^2 + 0.6x^2-y^2}}$ orbital in Co(CO)$_2$/Cu(001), which confirms this assumption.

\subsection{Structure dependence of the local moment in  Co(CO)$_2$/Cu(001)}\label{sec:lolcmomdicarb}
\begin{figure}[H]
  \centering
  \includegraphics[width=0.7\textwidth]{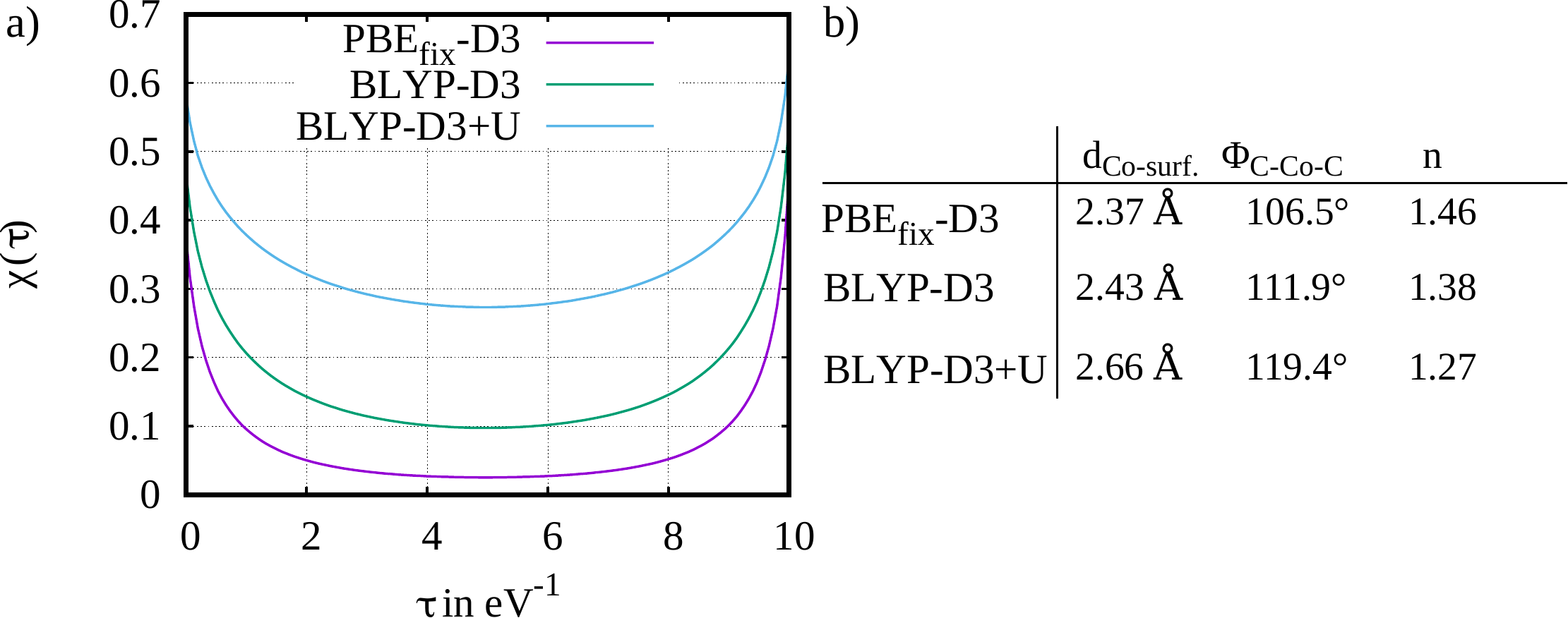}
  \caption{a) Spin--spin correlation $\chi(\tau)$ as obtained from PBE++ at $\beta$ = 10~eV$^{-1}$ ($U$ = 4.0~eV and $J$ = 0.9~eV) of the Kondo-relevant orbital $3d_{\mathrm{0.8z^2 + 0.6x^2-y^2}}$ in Co(CO)$_2$/Cu(001) for different optimized structures. b) Co-surface distance $d_{\mathrm{Co-surf.}}$, C-Co-C bonding angle $\phi_{\mathrm{C-Co-C}}$ and $3d_{\mathrm{0.8z^2 + 0.6x^2-y^2}}$ filling $n$ (PBE++) for Co(CO)$_2$/Cu(001) as obtained from different optimized structures.}
  \label{dicarb:struc_sus_10}
 \end{figure}

In the context of the preceding section, one could ask how strongly the results for Co(CO)$_2$/Cu(100) would change for a different molecular structure (see Section \ref{struc} for more details about how strongly structural parameters can vary depending on the computational parameters), as this might allow for a deeper insight into structure--property relations for the Kondo effect. 
To investigate the dependence of the magnetization on the adsorption distance,  we show  the spin--spin correlation function  at $T$ = 1160~K of the Co $3d_{\mathrm{0.8z^2 + 0.6x^2-y^2}}$ orbital for selected structures  (Figure \ref{dicarb:struc_sus_10}). For the BLYP-D3 and PBE$_{\mathrm{fix}}$-D3-optimized structures, $\chi(\tau)$ drops faster 
than for the BLYP-D3$+U$-optimized structure. This points to a stronger screening of the local moment if the adsorption distance is decreased. At the same time the adsorption distance is increased, the bonding angle $\phi_{\mathrm{C-Co-C}}$ increases, too, and one could expect that the screening also depends on this angle. However, for the structures reported here, we can exclude that the spin--spin correlation function of the Co $3d_{\mathrm{0.8z^2 + 0.6x^2-y^2}}$ orbital is significantly affected by $\phi_{\mathrm{C-Co-C}}$ (see Supplementary Material for further details).

The occupation of the Co $3d_{\mathrm{0.8z^2 + 0.6x^2-y^2}}$ orbital shows a delicate dependence on the  structure,  similar to its dependence on the DC value as shown for the BLYP-D3$+U$ optimized structure. For this reason, it can be assumed that the transition to the Fermi-liquid regime  is similarly affected by changes in the adsorption distance and the  C-Co-C bonding angle, and therefore is very sensitive to the computational parameters with which the molecular adsorbates have been optimized. The structures under study here are likely a particularly challenging case because of their large structural flexibility (as opposed to more rigid phthalocyanines) and because of the challenges associated with describing CO binding to metal surfaces (``CO-puzzle''), as discussed in Sec.  \ref{struc}. 
These results also suggest that the Kondo effect in Co(CO)$_2$/Cu(001)  might be  controllable via external stimuli affecting the adsorption distance (in particular, interactions between STM tips and adsorbates~\cite{hein13,park10,lu18}). 

\section{DFT-based analysis of the Kondo effect: What can we learn without solving the Anderson impurity model?}\label{sec:analysis}
 \begin{figure*}
  \centering
  \includegraphics[width=1.0\textwidth]{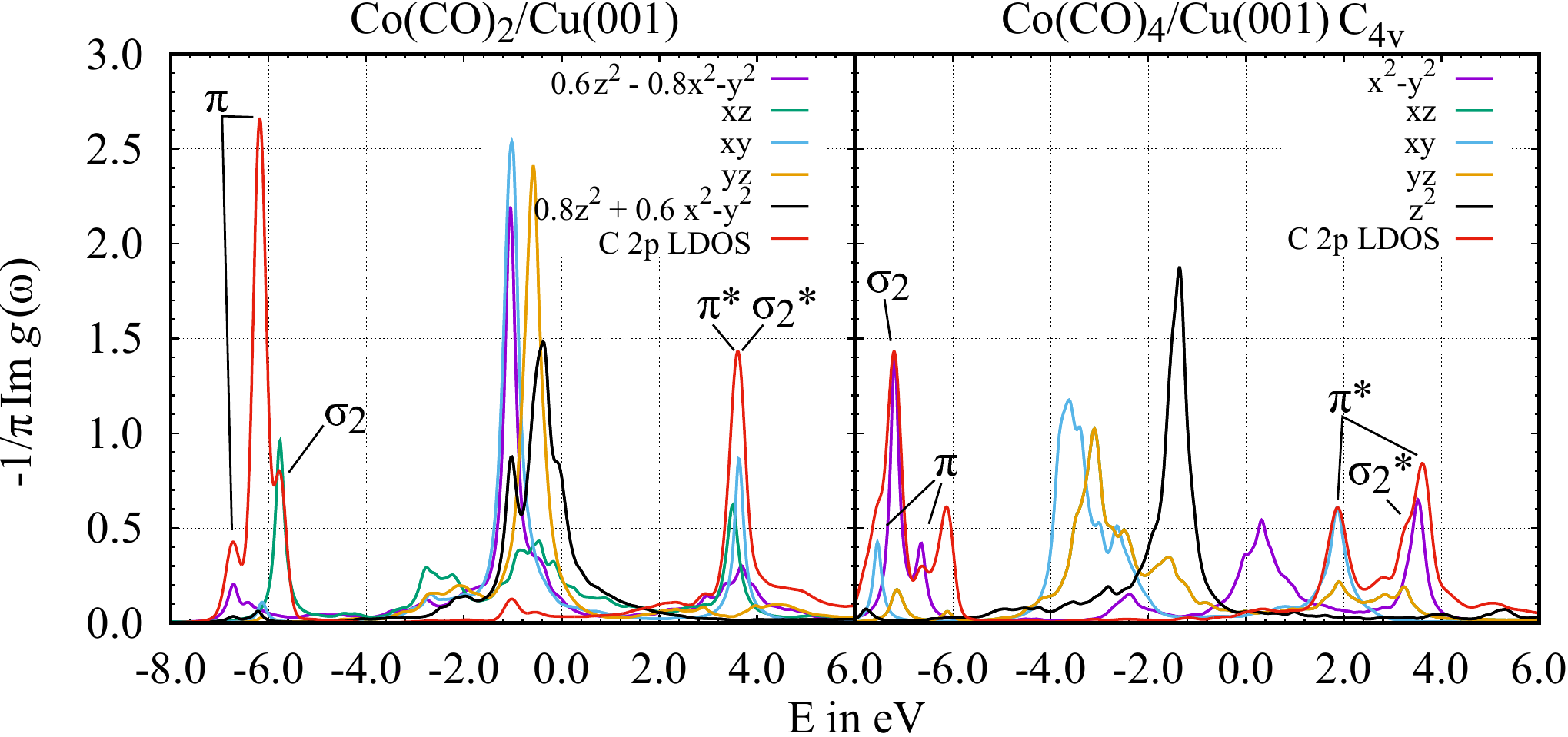}
  \caption{Projected density of states of the Co $3d$ orbitals (-$\frac{1}{\pi}$Im $g_i(\omega)$), and local density of states of the $2p$ orbitals  (sum over all $2p$ orbitals) of one of the C atoms. Results obtained from PBE (based on the BLYP-D3$+U$ optimized structures).}
  \label{carb_pdose}
 \end{figure*}
To gain further insight into the Kondo properties as obtained from solving the Anderson impurity model, it can be helpful to consider the spin-unpolarized PBE electronic structures, as they were  used  to parametrize the AIM (based on the BLYP-D3$+U$ optimized structures). 



In Figure \ref{carb_pdose}, we show the projected density of states ($-\frac{1}{\pi}\mathrm{Im} g_i(\omega)$) of the Co $3d$ orbitals, as well as the  C $2p$ local density of states (LDOS), which is the  sum of the projected density of states of the C $2p$ orbitals. They should exhibit peaks at the $\sigma_2/\sigma_2^*$ and $\pi/\pi^*$ orbitals of the CO ligands (a schematic representation of the molecular orbital diagram of an isolated CO molecule is provided in the Supplementary Material). From this, one can learn which of the Co $3d$ orbitals are interacting with the CO ligands, indicated by features in the Co $3d$ PDOS at the same position as the C $2p$ LDOS. This information might be useful to learn more about  chemical control of the Kondo effect by increasing the number of CO ligands, as this will affect the coupling of the Co $3d$ orbitals with the rest of the system.

For Co(CO)$_2$/Cu(001), the $3d_{\mathrm{0.8z^2+0.6x^-y^2}}$ orbital shows only small features at the C $2p$ LDOS, as a consequence of this orbital interacting only little with the CO ligands (as, e.g., compared to the  $3d_{\mathrm{xz}}$ orbital). In contrast, the $3d_{\mathrm{x^2-y^2}}$ orbital in Co(CO)$_4$ ($C_{\mathrm{4v}}$) interacts with the CO ligands, as indicated by the features in the PDOS at the position of the $\sigma_2/\sigma_2^*$ orbitals of CO. This coupling might increase the Kondo temperature, as discussed below in more detail. 

\begin{figure*}
  \centering
  \includegraphics[width=1.0\textwidth]{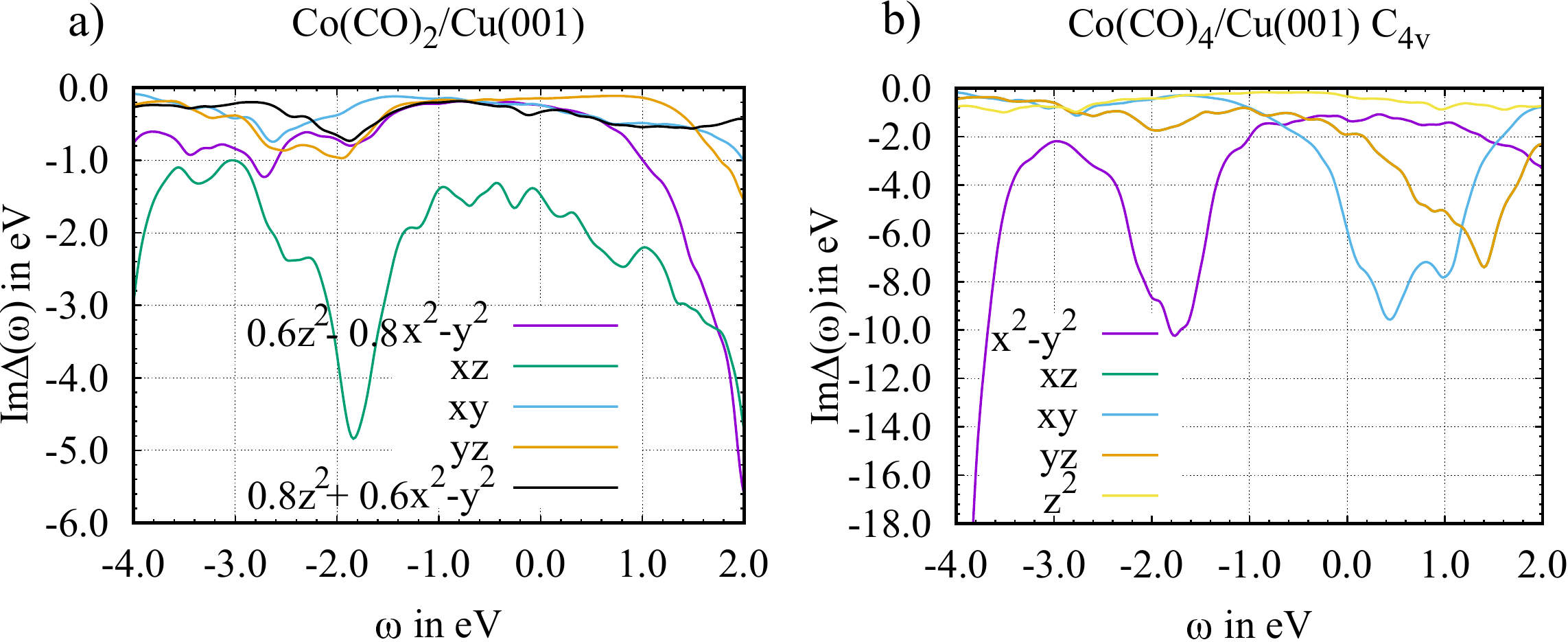}
  \caption{Imaginary part of the hybridization function  of a) Co(CO)$_2$ and b) Co(CO)$_4$ ($C_{\mathrm{4v}}$) on Cu(100) as obtained form PBE (based on the BLYP-D3$+U$ optimized structures).}
  \label{carbonyls_hyb}
 \end{figure*}

Considering the value of the energy-dependent hybridization function (Figure \ref{carbonyls_hyb}) at the Fermi energy ($\omega=0.0$~eV)), one gets a more complete picture of how strongly the Co $3d$ orbitals are coupled with the rest of the system. This value is known from the simplest Kondo model (one-band with a constant hybridization)\cite{hewson} to be directly connected to the Kondo temperature.

Focusing on the Kondo relevant orbitals of Co(CO)$_2$/Cu(001) ($3d_{\mathrm{0.8z^2+0.6x^-y^2}}$) and Co(CO)$_4$/Cu(001) ($3d_{\mathrm{x^2-y^2}}$), the energy-dependent hybridization function is in both cases rather featureless in the range of $\omega$ = -1.0~eV to $\omega$ = +1.0~eV. As shown in Table \ref{hyb0}, the value at $\omega$ = 0.0~eV for the $3d_{\mathrm{x^2-y^2}}$ orbital of Co(CO)$_4$/Cu(001) is about four times larger than for the $3d_{\mathrm{0.8z^2+0.6x^-y^2}}$ orbital in Co(CO)$_2$/Cu(001). This supports our assumption that the interaction of the $3d_{\mathrm{x^2-y^2}}$ orbital of Co(CO)$_4$ ($C_{\mathrm{4v}}$) with the CO $\sigma_2/\sigma_2^*$ orbitals increases the coupling at the Fermi energy, which in turns results in a larger Kondo temperature ($T_K  \approx$ 165 K) as discussed in Sec. \ref{sec:kondo}. Assuming that the Kondo temperature for Co(CO)$_2$/Cu(001) is lowered by the same factor as the hybridization at the Fermi energy of the $3d_{\mathrm{0.8z^2+0.6x^-y^2}}$ orbital, one would expect it to be around $T_{\mathrm{K}} \approx$ 41~K. The lowest electronic temperature under consideration in the DFT++ calculations discussed in Sec. \ref{sec:kondo} was $T$ = 46.4~K, which could explain why within the fully localized limit, we were not able to reach the transition to the Fermi liquid regime. According to Sec. \ref{sec:lolcmomdicarb}, the decreased adsorption distance of Co(CO)$_2$/Cu(001) in, e.g., the BLYP-D3-optimized structure, shows a hybridization of the Kondo-relevant orbital at $\omega$ = 0.0~eV of 0.52~eV. Compared with  Im$\Delta(0~\mathrm{eV})$ = 0.33~eV obtained for the BLYP-D3$+U$ structure, the Kondo temperature would therefore probably shift towards larger values for this structure. 

  \begin{table}[h]
     \caption{Hybridization in eV at the Fermi energy -Im$\Delta(0~\mathrm{eV})$ as obtained from spin-unpolarized PBE (based on the BLYP-D3$+U$ optimized structures) for different carbonyl cobalt complexes on Cu(001). For Co(CO)$_2$, the orbitals are labeled according to their largest contribution after diagonalizing the Co $3d$ sub-space.}
    \centering
    \begin{tabular}{c c c c c  c}
   \toprule[1.5pt]
   		& $d_{\mathrm{x^2-y^2}}$	& $d_{\mathrm{xz}}$	& $d_{\mathrm{xy}}$	& $d_{\mathrm{yz}}$	& $d_{\mathrm{z^2}}$\\
   \midrule[0.75pt]
   Co(CO)$_2$   &   0.24	 &  1.48   &  0.24	 &  0.1   &  0.33	\\
   Co(CO)$_4$ ($C_{\mathrm{4v}}$)  &   1.31  &  1.91   &  5.87   &  1.91  &  0.33   \\  
   \bottomrule[1.5pt]
    
    \end{tabular}
    \label{hyb0}
    \end{table}
 
%
  \section{Conclusion}\label{concl}
%
Chemical and mechanical control of the Kondo effect in molecular adsorbates is an intriguing subject, which promises insight into strong electron correlation. We have studied such control at the example of experimentally characterized cobalt carbonyl complexes on Cu(001) from a theoretical point of view, employing both DFT++ approaches for a full description of correlation, and DFT-derived properties for a conceptual understanding of structure--property trends. We find that it is indeed possible to optimize structures with DFT (employing BLYP-D3+U) whose Fermi liquid properties are compatible with the experimentally observed trend of larger Kondo temperatures with increasing number of ligands for the di- and tetracarbonyl complex (constraining the latter to $C_{4v}$ symmetry). We can trace back this behavior to an increased hybridization at the Fermi energy, which correlates with a stronger interaction of the Kondo-relevant $3d$ orbital with the CO ligands for the tetracarbonyl. This Kondo-relevant orbital is the $d_{x^2-y^2}$ in both cases, with a strong admixture of $d_{z^2}$ for the dicarbonyl system. It would be interesting to compare these data with newly developed approaches, in which a general projection scheme allows for extending the correlated impurity from the cobalt $3d$ orbitals to molecular orbitals which include part of the CO ligands~\cite{Droghetti2017,schu18}. 

Our data also point to the challenges such systems pose for present-day first-principles electronic structure methods: 
The structural flexibility of cobalt carbonyl complexes, along with the known difficulty of describing direct carbonyl--metal binding by present-day DFT, implies that predictive modeling of their Kondo properties is virtually impossible. In particular, no atomistic structure could be obtained for the tricarbonyl which is compatible with the experimentally observed Kondo effect (and with the lack of threefold symmetry suggested by STM data). Furthermore, all DFT protocols employed here suggest that for the tetracarbonyl, a $C_{2v}$-symmetric structure is by at least 30 kJ/mol more stable than a $C_{4v}$-symmetric one, yet only for the latter can we obtain Fermi liquid properties consistent with the experimentally observed Kondo effect. This suggests that the fourfold-symmetric structure observed in the STM results from the intrinsic symmetry of the molecule rather than from a rotational process. This discrepancy might result from the deficiencies of present-day DFT, or from a kinetic stabilization of the $C_{4v}$-symmetric structure in the experiment.  This is an example of employing spectroscopic data rather than solely total energies for identifying molecular structures compatible with the experiment, as also done, for example, in theoretical EXAFS studies~\cite{lube14}.
For systems with less pronounced structural flexibility, such as metal phthalocyanines, it is likely that available first-principles methods are more reliable at present. On the upside, the strong dependence of Kondo properties on structural parameters suggested by our data could imply that these Kondo properties can be controlled mechanically, in particular by interactions with an STM tip.

  \section{Supplementary Material}

See Supplementary Material for further details on atomistic structures, 
on spectral properties and spin--spin  correlation data of
Co(CO)$_3$ on Cu(001) and Co(CO)$_2$ on Cu(001) ($C_{\mathrm{2v}}$),
on molecular orbitals of the CO molecule, and on the angular dependence 
of spin--spin correlation and hybridization function for Co(CO)$_2$ on Cu(001).

  \section{Acknowledgment}
The authors acknowledge the high-performance-computing team of the Regional High-Performance Computing Center at  University of Hamburg and the North-German Supercomputing Alliance (HLRN) for technical support and computational resources, and the DFG for financial support via SFB 668.
  \section{Appendix}\label{appendix}

 \subsection{The effect of the CO--surface interaction on the Co $3d$ hybridization function}
  \begin{figure}[H]
   \centering
   \includegraphics[width=0.5\textwidth]{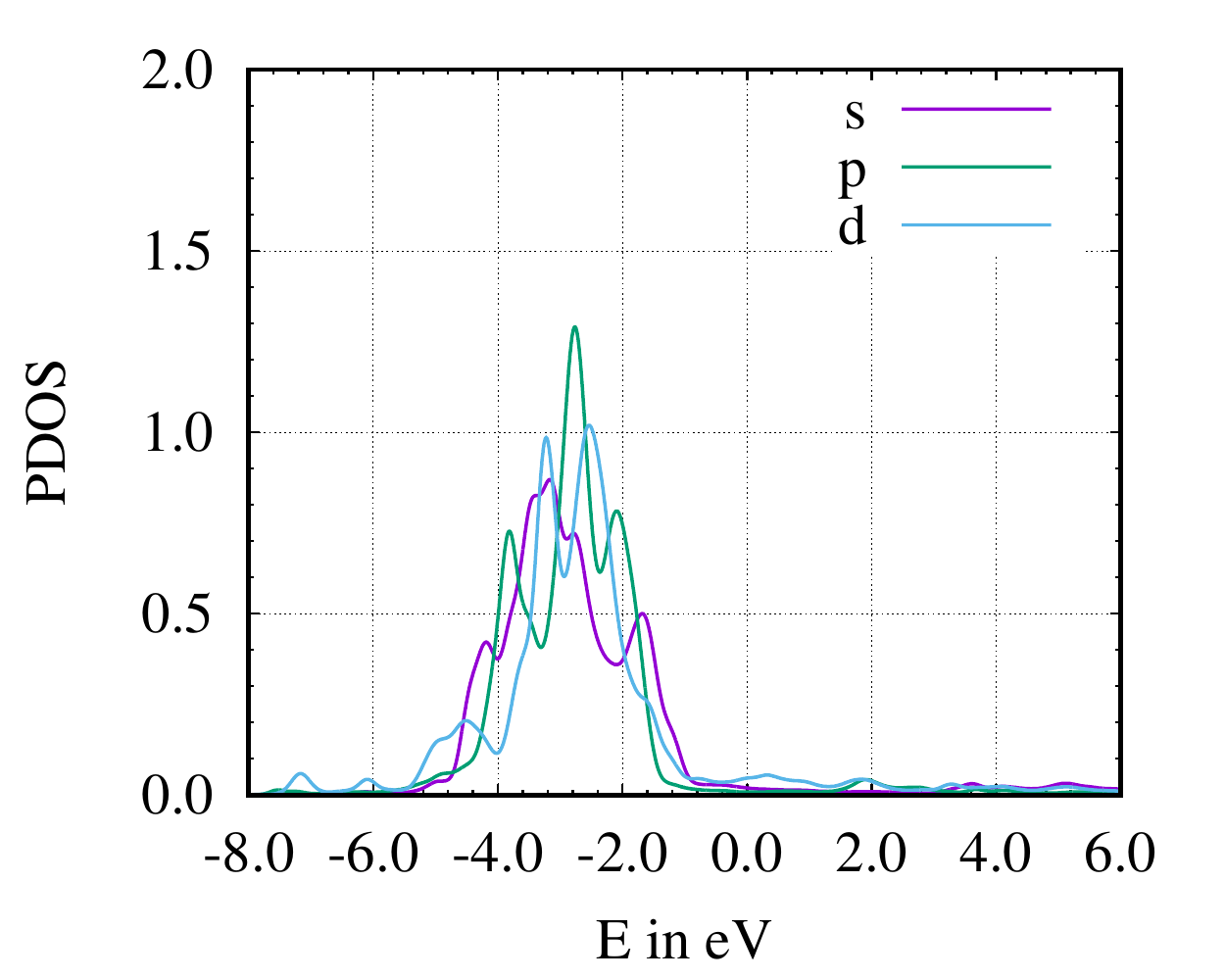}
   \caption{Density of states of the $4s$, $4p$ and $3d$ orbitals of one of the Cu surface atoms. Results obtained from PBE.}
   \label{Cu_tetrac4_dos}
  \end{figure}
\begin{figure}[H]
  \centering
  \includegraphics[width=1\textwidth]{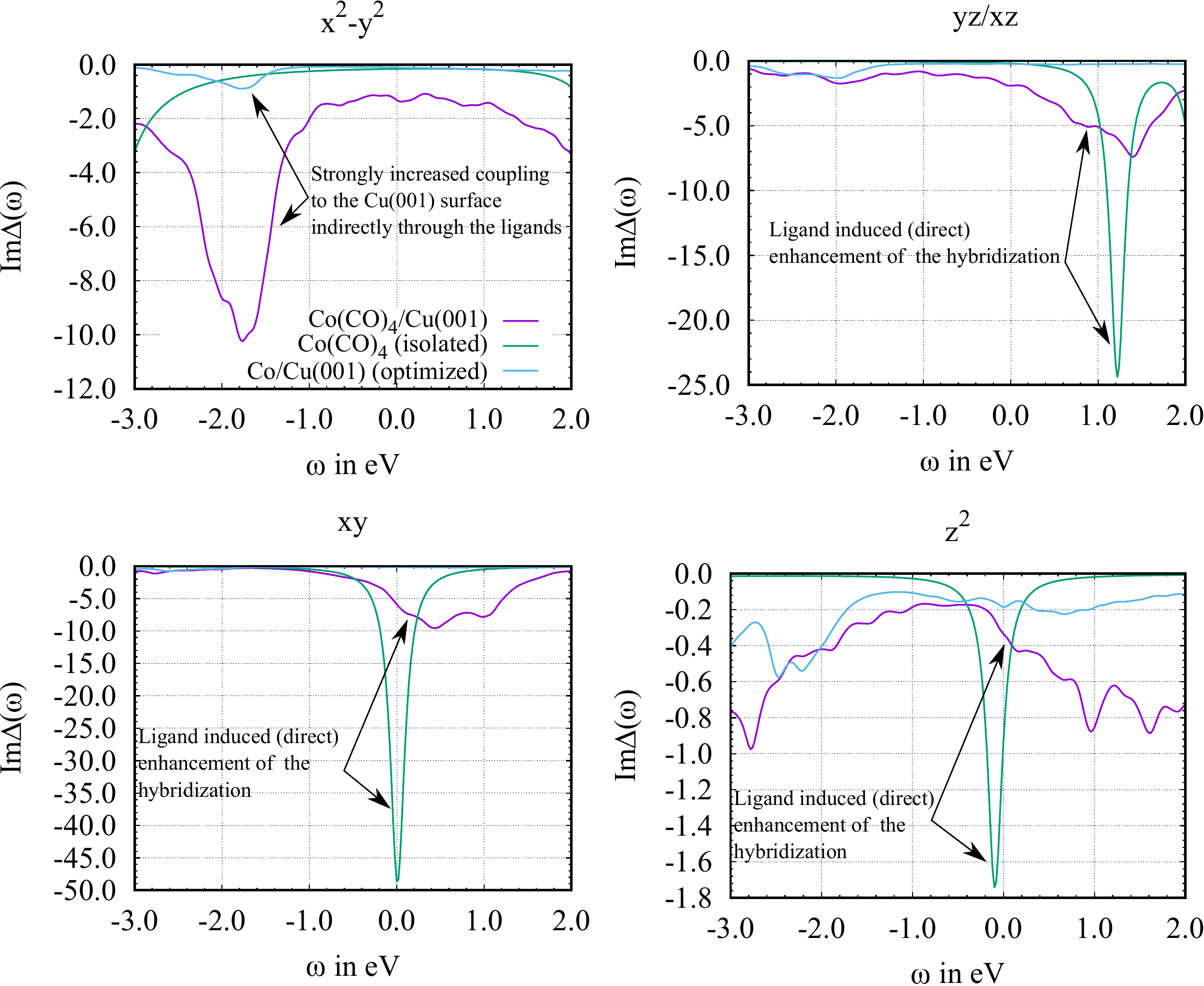}
  \caption{Co $3d$ hybridization functions as obtained from an isolated (optimized) Co on Cu(001), from an isolated Co(CO)$_4$ molecule as obtained by removing the surface atoms from the optimized Co(CO)$_4$/Cu(001), and from Co(CO)$_4$ on Cu(001). Hybridization functions as  obtained from PBE, based on BLYP-D3+$U$ optimized structures.}
  \label{hyb_analysis_tetrac4}
 \end{figure}

It would be interesting to study the contribution to the hybridization function of the Co $3d$ orbitals coming from the CO ligands directly, or indirectly as caused by the ligands being coupled to the Cu(001) surface. We try to make a step towards answering this question at the example of Co(CO)$_4$/Cu(001) in $C_{\mathrm{4v}}$ symmetry. For this purpose, we compare the hybridization functions in Figure \ref{hyb_analysis_tetrac4} of all Co $3d$ orbitals as obtained from an isolated Co on Cu(001), for an isolated Co(CO)$_4$ molecule (no surface), and for Co(CO)$_4$ on Cu(001). For the isolated molecule, we start from the optimized system and removed all Cu atoms, in order to see the contribution to the hybridization function of the CO ligands only. While this analysis neglects effects of, e.g., the surface on the CO ligands, which may in turn affect the way these ligands contribute to the hybridization of the Co orbitals, we do expect an elucidating qualitative picture of the relative importance of ligands and surface. 

Note that the plots in Figure \ref{hyb_analysis_tetrac4} are differently scaled on the $y$-axis. The hybridization functions of all Co $3d$ orbitals for an isolated Co atom on Cu(001) are rather small and featureless. In all cases, however, there is a small bump at roughly $E$ = -1.8~eV to $E$ = -2.5~eV, which comes from the increased DOS of the Cu surface at this energy (as shown in Figure \ref{Cu_tetrac4_dos}).

Considering the Co $3d$ hybridization functions (Figure \ref{hyb_analysis_tetrac4}) of the isolated molecule should give an impression of the contribution of the ligands to the hybridization of the Co $3d$ orbitals in the full system.
For the isolated molecule, the hybridization functions of all $3d$ orbitals (except for the $d_{\mathrm{x^2-y^2}}$ orbital)  exhibit a sharp feature close to the Fermi energy, which is the reason for the hybridization function of these orbitals  being increased in the vicinity of $E$ = 0.0~eV for the full system (Co(CO)$_4$/Cu(001))\footnote{For the $d_{\mathrm{z2}}$ orbital this effect is only small, as can be seen by the low intensity of the peak of $\Delta(\omega)$ in case of the isolated molecule,  in contrast to the $d_{\mathrm{xy}}$ and $d_{\mathrm{xz/yz}}$ orbitals.} compared to an isolated Co atom on Cu(001). We believe that this enhancement is directly induced by the hybridization with the ligands. For the $d_{\mathrm{x^2-y^2}}$ orbital the  hybridization at $E$ = -1.8~eV is significantly increased  compared to Co/Cu(001), although for the isolated molecule we do not observe a peak in $\Delta(\omega)$  at this energy. Thus, we conclude that the increased hybridization is indirectly caused by the CO ligands, as they not only  couple strongly to the Co  $d_{\mathrm{x^2-y^2}}$ orbital (as pointed out in Section \ref{sec:analysis}), but also to the Cu(001) surface.

\end{document}